\documentclass[conference]{IEEEtran}
\usepackage{cite}

\ifCLASSINFOpdf
   \usepackage[pdftex]{graphicx}
\else
   \usepackage[dvips]{graphicx}
   \DeclareGraphicsExtensions{.eps}
\fi
\usepackage{amsmath}
\ifCLASSOPTIONcompsoc
  \usepackage[caption=false,font=normalsize,labelfont=sf,textfont=sf]{subfig}
\else
  \usepackage[caption=false,font=footnotesize]{subfig}
\fi

\usepackage{stfloats}

\usepackage{multirow}

\newtheorem{proposition}{Proposition}

\newtheorem{remark}{Remark}
\begin{document}
	\title{Mitigating primary emulation attacks in multi-channel cognitive radio networks: A surveillance game}	
	\author{
		\IEEEauthorblockN{Duc-Tuyen Ta$^1$, Nhan Nguyen-Thanh$^1$, Patrick Maille$^2$, Philippe Ciblat$^1$, and Van-Tam Nguyen$^{1,3}$}		
		\IEEEauthorblockA{$^1$LTCI, CNRS, Télécom ParisTech, Université Paris Saclay, 75013, Paris, France\\
			$^2$ Telecom Bretagne/IRISA, Institut Mines-Telecom, Rennes, France\\
			$^3$Department of EECS, University of California at Berkeley, USA}
	}
	\maketitle
	\begin{abstract}
		\boldmath
		Primary User Emulation Attack (PUEA), in which attackers emulate primary user signals causing  restriction of secondary access on the attacked channels, is a serious security problem in Cognitive Radio Networks (CRNs). An user performing a PUEA for selfishly occupying more channels is called a selfish PUEA attacker. Network managers could adopt a surveillance process on disallowed channels for identifying illegal channel occupation of selfish PUEA attackers and hence mitigating selfish PUEA. Determining surveillance strategies, particularly in multi-channel context, is necessary for ensuring network operation fairness. In this paper, we formulate a game, called \textit{multi-channel surveillance game}, between the selfish attack and the surveillance process in multi-channel CRNs. The \emph{sequence-form representation} method is adopted to determine the Nash Equilibrium (NE) of the game. We show that performing the obtained NE surveillance strategy significantly \textit{mitigates} selfish PUEA.
	\end{abstract}
	\IEEEpeerreviewmaketitle
\section{Introduction}
	
	Cognitive radio (CR) is introduced as a solution to improve spectrum utilization by enabling secondary access to licensed spectrum. Discovering spectrum holes is therefore essential. To explore the spectrum opportunities, there are two approaches: spectrum sensing and database-driven. In the spectrum sensing approach, the primary users' activity is explored by measuring the spectrum environment, while in the database-driven approach, the information of spectrum usage for CR users is provided by a database server. Compared to the database-driven approach, the spectrum sensing approach is cheaper and more flexible for a wide range of networks~\cite{Murty2011}.  However, spectrum sensing faces a serious security risk: Primary User Emulation Attack (PUEA)~\cite{Ruiliang2008_PUE,Ruiliang2008_both, Mohapatra2013_PUE,6231141,Li2010_PUE,7103339,7194088}. In PUEA, the attacker transmits an emulated Primary User (PU) signal which could lead to disallowed state on the attacked channels (i.e., the channels which spectrum sensing system claims to be busy after the sensing period). The attacker therefore gains exclusive spectrum right.
	Depending on goals, PUEA can be categorized into two types: selfish and malicious. A malicious PUEA targets at ruining the operation of networks, hence it is similar to the jamming attack or the denial-of-service attack. A selfish PUEA, however, aims at selfishly occupying the attacked channel for data transmission. In that sense, selfish PUEA is associated with an illegal benefit degrading the fairness of CR Networks (CRNs) and possible interference threating to primary systems. In this study, we therefore focus on how to mitigate the selfish PUEA on CNRs.

	Several earlier works have investigated the issues of selfish PUEA. Most of them adopt detection methods based on additional information, such as the locations of both primary and secondary users~\cite{Ruiliang2008_PUE} or the frequency deviation feature of FM signal~\cite{Mohapatra2013_PUE} to identify the PUEA attackers. However, those methods are only applicable on special cases where added information, or the imperfection of attacked signals is available. The other works applied the game-theoretical framework to analyze the risk of PUEA~\cite{Felegyhazi2006,6231141,bwang2010, Li2010_PUE,7103339, 7194088} since there are the opposing objectives between the attacker and the network manager.  In~\cite{6231141}, the authors formulate a non-cooperative multistage game between a selfish PUEA attacker and a secondary node on the data transmission phase. A dogfight spectrum game between a PUEA attacker and a CR user is formulated in~\cite{Li2010_PUE}. In that work, PUEA attacking signals are treated as jamming signals and channel hopping is proposed as a solution for mitigating PUEAs. However, there is still vulnerability if the attacker conducts multiple channel attacks.
	
	A successful selfish PUEA is usually followed by selfishly using of the attacked channel by the attacker. Therefore, it is possible to determine the illegal accessing in any communication link through the user’s identification. In our previous works~\cite{7103339,7194088}, we have proposed a surveillance process to mitigate the influence of selfish PUEA by monitoring data traffic at the beginning of the data frame on an occupied channel. To determine the surveillance strategy of the network manager, we use a game-theoretic approach to formulate the relation between the attack and the surveillance process in a \emph{single} channel. The best strategies of the attacker and the network manager are figured out in closed-form, as a Nash equilibrium (NE) point. However, CRNs usually work on multiple frequency bands, and because of the rapid expansion of software-defined radio, the attacker can launch multi-channel selfish PUEA. For such a case, a sequential monitoring plan can be used, however, at the cost of long surveillance time. In this paper, we therefore consider the multi-channel surveillance process to mitigate the influence of the selfish PUEA in CRNs.
	
	The multi-channel surveillance model is more \emph{complicated} but more \emph{realistic} than the single-channel model. We formulate the relation between the selfish PUEA and the surveillance process as a two-player game in extensive form and consider the Nash Equilibrium (NE) for the surveillance strategy. Note that this approach can be extended to mitigate the influence of malicious PUEA or unknown-attacking PUEA in CRNs.
	
	Typically, the network manager observes the attacker's action only indirectly, through the sensing results. Hence, the formulated game is an incomplete and imperfect information game. Finding a Nash equilibrium solution in this game is more complicated than in perfect-information games. We employ the \emph{sequence-form representation} method~\cite{koller1996efficient, von1996tracing} instead of the conventional (benchmark) \emph{strategic-form representation} method~\cite{Harsanyi2004} to determine the best strategy for the defender and the attacker. We prove that the sequence-form representation is much more efficient than the strategic-form representation method. We then analyze and interpret the impact of the system parameters includes the PU's presence probabilities, the network demand, as well as the penalty factor on the obtained NE strategies.
\section{System Model}
We call the \emph{attacker}, the representative of  PUEA users and the \emph{defender}, the resource manager of the network which monitors the traffic on disallowed channels to detect selfish attackers. Let $N$ be the number of available channels, $M$ be the maximum number of channels that the attacker can attack and $L$ be the maximum number of channels that the defender can monitor.
	
	The timing frame for network operation is the same as our previous work (Figure 1 in ~\cite{7103339}). Before the transmission (\emph{data} phase), a \emph{sensing} phase, possibly followed by a \emph{surveillance} phase, is carried out. For each channel, because of the non-ideal sensors, two possible sensing results could be obtained: ``\textit{allowed}", and ``\textit{disallowed}". An attacker can implement a selfish PUEA by transmitting the emulated primary user signals during the sensing time. We assume the sensors cannot distinguish the emulated and authentic primary signals. Consequently, the PUEA will not be detected in the sensing time. During a PUEA, the attacker cannot know the true status of the primary user on the attacked channel since it cannot sense for PU signal at the same time. This means the PUEA attacker conducts the attack in a blind condition regarding the primary signal status. Considering the defense against selfish PUEA threats, we assume that a fixed format data frame is used for exchanging data with all CR users including selfish users. The format contains the identifying information of a user such as the \textit{medium access control} address. Therefore, CR users can be identified by observing transmitted signals in data time. Channel surveillance processes which are conducted on monitored secondary access channels can identify the selfish attackers if they are presented. Once a selfish attacker has been detected, punishments such as network isolation or bandwidth limitation can be adopted to penalize the attacker. This surveillance process is assumed to be implemented by the resource manager.
	
	Before the analysis, we summarize the specific notations that will be used throughout the paper (Table~\ref{tab:definition}).
	\vspace{-0.2cm}
	\begin{table}[t]
		\centering
		\caption{Notations (at $i$ channel)}
		\label{tab:definition}
		\begin{tabular}{lp{72mm}}
			Notation & Meaning\\
			\hline
			${\pi_{i}}$ & The presence probability of PU. \\
			${p_N^{i}}$ & The probability that the sensor answers disallowed channel  when the attacker does not attack. \\
			${p_A^{i}}$ & The probability that sensor answers disallowed channel when the attacker attacks.\\
			${\rho_N^{i}}$ & The probability that the channel is not used by the PU while the sensor claims disallowed and the attacker does not attack. \\
			${\rho_A^{i}}$ & The probability that the channel is not used by the PU while the sensor claims disallowed and the attacker attacks.\\
			$C^{i}_A$ & The implementing cost of the selfish PUEA.\\
			$G^{i}_A$ & The using gain of selfish PUEA attacker at one data frame.\\
			$C^{i}_S$ & The monitoring cost of the resource manager at the channel.\\		
			$G^{i}_M$ & The capturing gain for detecting illegal attack during the surveillance process of data frame.\\
			$P^{i}$ & The penalty value for being captured at the channel.\\
			\hline
		\end{tabular}
	\end{table}	
\section{Game Formulation}	
\begin{figure}[!t]
		\centering
		\includegraphics[width=1\linewidth]{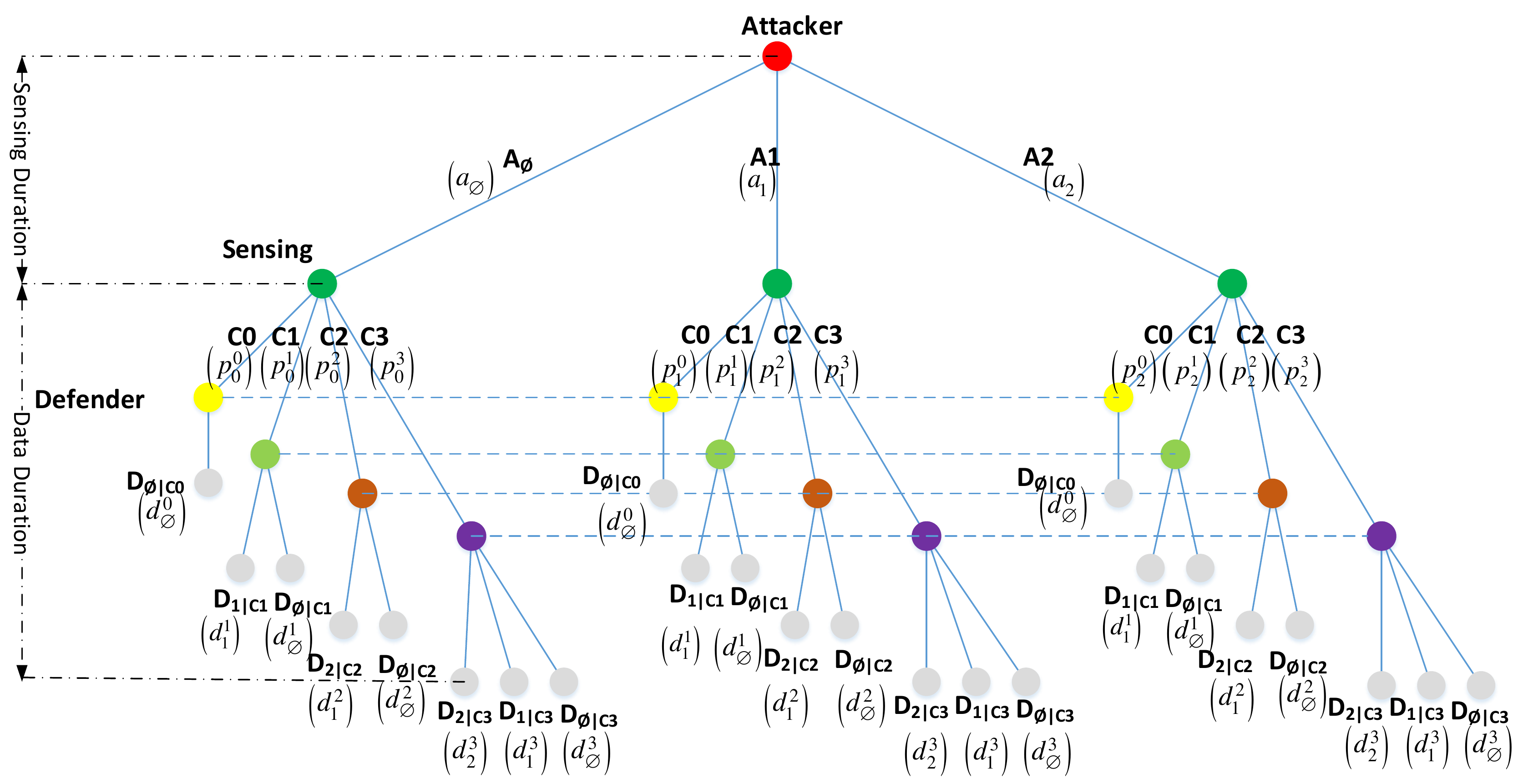}
		\caption{Two-channel surveillance game}
		\label{fig:model}
	\end{figure}

	We formulate a two-player game in extensive form, called the \textit{multi-channel surveillance game} (MSG), to present the relationship between the multi-channel surveillance process and the selfish PUEA in CRNs. There are two players: \textbf{Attacker} (player 1) who represents selfish PUEA attackers, and \textbf{Defender} (player 2) who represents the resource manager. Figure~\ref{fig:model} illustrates the MSG game for a CRN with two channels ($N=2$), the attack capability of the attacker $M=1$ and the monitoring capability of the defender $L=1$.
	\subsubsection{Strategies}
	We denote by  ${A_\emptyset }, {A_1},{A_2}, \ldots ,{A_{K_1}}$ the actions of the attacker, where ${A_\emptyset }$ corresponds to the action \textbf{No Attack} and ${A_i}$ corresponds to the action \textbf{Attack} of a non-empty subset of available channels. The attacker has ${K_1}$ actions leading to its pure strategy set $\Sigma_A$.
	\begin{equation}
	\Sigma_A = \{{A_\emptyset }, A_1, A_2, \ldots, A_{K_1}\}
	\end{equation}
	where $K_1=\sum_{i=1}^{\min(M,N)} {{N}\choose{i}}$.
	
	Let $\bf{C}$ be the set of the sensing result of all channels and $\bf{C}$ includes $2^N$ elements. For a sensing result ${C_k \in {\bf{C}}}$ ${({k=0, \ldots, 2^{N}-1})}$, let $|C_k|$ be the number of the disallowed channels. We denote by ${{D_{\emptyset|{C_k}}},D_{{1}|{C_k}},D_{{2}|{C_k}}, \ldots ,D_{{|C_k|}|{C_k}}}$ the possible actions of the defender in which ${D_{\emptyset|{C_k}}}$ corresponds to the \textbf{No Surveillance} action and $D_{{j}|{c_k}}$ corresponds to the \textbf{Surveillance} action of a disallowed channel subset belong $C_k$. The pure strategy set of the defender has $K_2$ actions leading to its pure strategy set $\Sigma_D$.
	\begin{equation}
	\Sigma_D = \bigcup\limits_{k=0}^{2^N-1}  \{D_{\emptyset|{C_k}},D_{{1}|{C_k}},D_{{2}|{C_k}}, \ldots ,D_{{|C_k|}|{C_k}}\}_{C_k \in {\bf{C}}},
	\end{equation}
	and ${K_2} = 1 + \sum\limits_{m = 1}^{\min(N,L)} {\sum\limits_{n = 1}^m {{m}\choose{n}} \times {{N}\choose{m}} } $.
	
We denote by $\Delta_A$ the mixed strategy set of the attacker and $\Delta_D$ the mixed strategy set of the defender.
\subsubsection{Expected payoff}	
After sensing phase, there are two possible sensing results for $t$ channel: 1) disallowed and 2) allowed. Depending on the action of two players, their payoffs at $t$ channel are presented in Table~\ref{tab:payoff}. These payoffs are computed by considering the present of PU which is represented by the values of ${\rho_A^{i}}$  and ${\rho_N^{i}}$. 
	\begin{table}[hbt]
		\centering
		\caption{Payoffs of the attacker and the defender at $t$ channel} 
		\label{tab:payoff}
		\begin{tabular}{p{1.1cm}|p{2.1cm}|p{1.4cm}|p{2.5cm}}
			& [\textit{Defender};\textit{Attacker}] & No Attack           & Attack                                                               \\
			\hline
			\multirow{2}{*}{Disallowed} & No Surveillance                              & {[}$0$;$0${]}        & {[}$0$;$-C^{t}_A+\rho^{t}_{A} G^{t}_A${]}                            \\
			\cline{2-4}
			
			& Surveillance                                  & {[}$-C^{t}_S$;$0${]} & {[}$-C^{t}_S+\rho^{t}_{A} G^{t}_S$; $-C^{t}_A-\rho^{t}_{A} P^{t}${]} \\
			\hline	 
			\multirow{1}{*}{Allowed} & No Surveillance                              & {[}$0$;$0${]}        & {[}$0$;$-C^{t}_A${]}                                                 \\		 
			\hline                                        
		\end{tabular}
	\end{table}			
	
We denote by $U_A^{t, D_{k|C_j}}$ the obtained payoff of the attacker by performing the Attack of $t$ channel while the defender plays $D_{k|C_j}$. Similarly, we denote by $U_D^{A_i, t}$ the obtained payoff of the defender by monitoring $t$ channel while the attacker plays $A_i$. For the action pair $\{A_i,D_{j|C_k}\}$, the payoffs of the defender ${U_D^{A_i, D_{k|C_j}}}$ and the attacker ${U_A^{A_i, D_{k|C_j}}}$ are given by
	\begin{eqnarray}
	U_A^{A_i, D_{k|C_j}}&=\sum_{t \in A_i} 	U_A^{t, D_{k|C_j}}
	\label{eq:att_pay}
	\\
	U_D^{A_i, D_{k|C_j}}&=\sum_{t \in D_{k|C_j}} 	U_D^{A_i, t}
	\label{eq:def_pay}
	\end{eqnarray}
	%
	
	We denote by $\Omega_A$ the expected payoff of the attacker and $\Omega_D$ the expected payoff of the defender, respectively. We have
	\begin{eqnarray}
	\Omega_D=\sum\limits_{i =1}^{K_1} \delta _a\left( A_i \right) \sum_{j=0}^{2^N-1} p_{C_j|A_i} \sum_{k=1}^{|C_j|} \delta _d( D_{k|C_j}) U_D^{A_i, D_{k|C_j}}\\
	\Omega_A=\sum\limits_{i =1}^{K_1} \delta _a\left( A_i \right) \sum_{j=0}^{2^N-1} p_{C_j|A_i} \sum_{k=1}^{|C_j|} \delta _d( D_{k|C_j}) U_A^{A_i, D_{k|C_j}}
	\end{eqnarray}
	where $p_{C_j|A_i}$	is the probability of the sensing result $C_j$ under the attacker's action $A_i$ and $\delta _a\left( A_i \right) \in \Delta_A$ and $\delta _d( D_{k|C_j}) \in \Delta_D$ are the mixed strategy of action $A_i$ and $ D_{k|C_j}$, respectively.
	\section{Nash equilibrium}
	For the MSG game, we explore the Nash equilibrium (NE) point, i.e., the point where each player has selected a best response (BR) strategy to other players' strategies. The BR are the strategies on which the player gains the highest payoff given other players' strategies. A NE strategy may be a "pure" or a "mixed" strategy.
	To determine the NE point, two approaches are considered: 1) the conventional \emph{strategic-form representation} and 2) the \emph{sequence-form representation}.	
	\subsection{Strategic-form representation}
	We first consider the conventional strategic-form representation, which is based on the Harsanyi transformation~\cite{Harsanyi2004} and the Lemke-Howson (L-H) algorithm~\cite{Lemke1964}. The Harsanyi transformation models all possible actions of a player, which are affected by the other players' actions and the nature choices. For the MSG game, however, the method results in an exponential increment in the size of the game. In particular, the size of payoff matrix in the MSG game adopting the strategic-form representation is $\left(K_1+1\right) \times K_3$, where $K_3$ is given by
	\begin{equation}
	{K_3} = \prod\limits_{k = 0}^{2^N-1} {{|C_k|}\choose{\min\left(M,|C_k|\right)}}
	\end{equation}
	where $ {N}\choose{k}$ denote a binomial coefficient indexed by $N$ and $k$.

	For the case that $M=L=1$, ${K_3={K^{*}_3} = \prod\limits_{k = 0}^N {{{\left( {k + 1} \right)}^{{N}\choose{k}}}}}$. It means that, the payoff matrix is $3 \times 12$ if $N=2$ and $5 \times (5 \times 12^{6})$ if $N=4$. It is significantly larger when $M$ and $L$ bigger than $1$. Consequently, it is very complicated to find the NE points of the game for the large $N$.
	\subsection{Sequence-form representation}
	In game theory, an \textit{extensive form} game includes the information about the sequencing of players' possible moves, the chance moves, payoffs for both players at the leaves and the information set at the decision nodes. If the game is \textit{perfect recall}, i.e. each player remembers its' earlier moves, each node has a unique path from the root. Such a game can be represented in the sequence-form where a \textit{sequence} is defined as a string listing the action choices of a particular player. In detail, for each node $h$ of player $i$, we define ${\sigma_h}$ as the sequence and $C_h$ as the set of choices of player $i$ at $h$. For each choice $c \in C_h$, the corresponding sequence of $i$ is $\sigma_h c$. Hence, the set of sequences ${\bf \Sigma}_i$ for player $i$ is given by
	\begin{equation}
	{\bf \Sigma}_i =\{\emptyset \} \cup \{ \sigma_h c |h \in H_i, c \in C_h\}
	\end{equation}
	where $H_i$ is the set of node of player $i$.
	
	Since the MSG game is a perfect recall game, we adopt the sequence-form representation to solve the game. \textit{The trick here is that we consider the sensing results as the elements of the attacker's sequence}. The sequence strategy set of the attacker then is
	\begin{align}
	{{\bf \Sigma}_{A}^{seq}} &= \left\{ {{\sigma _{Ai}},i = 1...K_4} \right\} \notag \\
	&= \left\{ \emptyset, A_\emptyset, A_1,\ldots, A_{K_1}, A_{\emptyset, C_0}, A_{\emptyset, C_{1}}, \ldots \right\}
	\end{align}
	where $K_4 = 1 + \left(K_1+1\right) + \left(K_1+1\right)\times 2^{N}$.
	
	The corresponding sequence strategy set of the defender is
	\begin{align}
	{{\bf \Sigma}_{D}^{seq}} &= \left\{ {{\sigma _{Dj}},j = 1...\left(K_2+1\right)} \right\} \notag \\
	&= \left\{ \emptyset,D_{\emptyset|{C_k}},D_{{1}|{C_k}}, \ldots ,D_{{|C_k|}|{C_k}}, \ldots
	\right\}_{C_k \in {\textbf{C}}}
	\end{align}
	The \textit{mixed strategy} for the sequence-form representation is presented by the probability of each sequence in the sequence strategy set. Let ${\Phi}_{A}$ and ${\Phi}_{D}$ denote the mixed sequence strategies for the attacker and the defender, respectively. From the definition of the sequence strategy set, we have
	\begin{align}
	{\Phi_{D}^{seq}} &= \{\phi^{i}_{d}\}_{i=1\ldots (K_2+1)}\\
	{\Phi_{A}^{seq}} &= \{\phi^{j}_{a}\}_{j=1\ldots K_4}
	\end{align}
	where $\phi^{i}_{a}$ is the probability of the attacker's $i^\text{th}$ sequence, $\phi^{j}_{d}$ is the probability of the defender's $j^\text{th}$ sequence.
	
	The relation between these mixed strategies is called the \textit{realization plan}. By default, for an empty sequence, the probability is $1$. For any node $h$, the mixed strategy of the sequence at $h$ is the sum of all mixed strategies from $h$. Therefore, we obtain the realization plans for the defender and the attacker:
	\begin{equation}
	\left\{ {\begin{array}{*{20}{l}}
		{{\phi_{d}\left(\emptyset\right) } = 1}\\
		{ \phi_{d}\left(D_{\emptyset|{C_k}}\right)+\sum_{i=1}^{|C_k|} \phi_{d} \left(D_{{i}|{C_k}}\right) = 1} \; \forall {C_k} \in \textbf{C}\\
		0 \leq \phi^{i}_{d} \leq 1, i=1\ldots K_2\\
		\end{array}} \right.
	\end{equation}
	and
	\begin{equation}
	\left\{ {\begin{array}{*{20}{l}}
		{{\phi_{a}\left(\emptyset\right) } = 1}\\
		{ \phi_{a} \left({A_{\emptyset}}\right)+\sum_{i=1}^{K_1} \phi_{a} \left(A_i\right) = 1}\\
		{\sum_{i=0}^{2^N-1} \phi_{a}\left(A_{\emptyset , C_i}\right)= \phi_a\left(A_\emptyset\right)}\\
		{\vdots} \\
		{\sum_{i=0}^{2^N-1} \phi_{a}\left(A_{K_1 , C_i}\right)= \phi_a\left(A_{K_1}\right)}\\
		0 \leq \phi^{i}_{a} \leq 1, i=1\ldots K_4\\
		\end{array}} \right
	.
	\end{equation}
	In general, these realization plans can be re-written using the following matrix form.
	\begin{equation}
	{\left\{ \begin{array}{*{20}{l}}
		{\bf{E}}{\Phi}_{A} = \bf{e}\\
		\Phi_A \ge 0
		\end{array} \right. } ,
	{\qquad \rm{and} \qquad}
	{\left\{ \begin{array}{*{20}{l}}
		{\bf{F}}{\Phi}_{D} = \bf{f}\\
		\Phi_D \ge 0
		\end{array} \right. } ,
	\end{equation}
	where ${\bf E}$ and ${\bf F}$ are called the constraint matrices, and $\bf{e}$ and $\bf{f}$ are vectors in which the first element is $1$, and the other elements are $0$.
	
	As defined above, each leaf of the game tree corresponds to a pair of sequences. Hence, for a pair of the sequence strategies $(\sigma_{Ai}, \sigma_{Dj})$, there are 3 cases: i) if a pair of sequences is ended at a leaf, the payoffs are computed by multiplying~(\ref{eq:att_pay}) and~(\ref{eq:def_pay}) with the corresponding probability of the change move leads to this leaf, ii) if a pair of sequences does not correspond to a leaf, the payoffs are zero and iii) if a pair of sequences corresponds to the leaves and consist the chance moves or the information set, the payoff is then the sum over all leaves that define the given pair of sequences.
	
	Let ${{\bf{\Pi }}_A}$ and ${{\bf{\Pi }}_D}$ denote the payoff matrix of the attacker and the defender in the sequence-form representation. The expected payoffs of the attacker $\Omega_A$ and the defender $\Omega_D$ therefore are computed by
	\begin{align}
	\Omega_A &= {\Phi}_{A}^{\bf{T}} {{{\bf{\Pi }}_A}{\Phi_D}} \\
	\Omega_D &= {\Phi}_{A}^{\bf{T}} {{{\bf{\Pi }}_D}{\Phi_D}}
	\end{align}
	An equilibrium is a pair $\left(\Phi_A,\Phi_D\right)$ of mutual best responses. In particular, if the realization plan $\Phi_D$ is fixed, then $\Phi_A$ is the best response to $\Phi_D$ if and only if it is an optimal solution of the linear program
	\begin{equation}
	\begin{array}{l}
	\mathop {{\rm{maximize}}} \limits_{\Phi_A} {\quad}{{\Phi}_{A}^{\bf{T}}}\left( {{{\bf{\Pi }}_A}{\Phi_D}} \right)\\
	{\text{subject to}}\quad {{\bf{E}}}{\Phi_A} = {{\bf{e}}}\\
	{\quad \qquad \qquad 0} \le {\Phi_A} \le 1
	\end{array}
	\end{equation}
	The dual form of linear program (21) is as follows.
	\begin{equation}
	\begin{array}{l}
	\mathop {{\rm{minimize}}}\limits_{\bf{p}} {\quad}{{\bf{e}}^T}{\bf{p}}\\
	{\text{subject to}} \quad {{\bf{E}}^T}{\bf{p}} \ge {{\bf{\Pi }}_A}{\Phi_D}\\
	{\quad \qquad \qquad}{{\Phi}_{A}^{\bf{T}}}\left( { - {{\bf{\Pi }}_A}{\Phi_D} + {{\bf{E}}^T}{\bf{p}}} \right) = 0
	\end{array}
	\end{equation}
	A similar program is established for the attacker strategy.
	
	In conclusion, the problem of finding a Nash equilibrium can be formulated as follows:
	\begin{equation}
	\begin{array}{l}
	\mathop {{\rm{minimize}}}\limits_{\bf{p}} {\quad}{{\bf{e}}^T}{\bf{p}} \\
	\mathop {{\rm{minimize}}}\limits_{\bf{q}} {\quad}{{\bf{f}}^T}{\bf{q}}\\	
	{\rm{subject \; to}} {\; \;} {{\bf{E}}^T}{\bf{p}} \ge {{\bf{\Pi }}_A}{\Phi_D}, 
	\quad {{\bf{F}}^T}{\bf{q}} \ge {\bf{\Pi }}_D^T{\Phi_D}\\
	{\qquad \qquad \quad}{{\Phi}_{A}^{\bf{T}}}\left( { - {{\bf{\Pi }}_A}{\Phi_D} + {{\bf{E}}^T}{\bf{p}}} \right) = 0\\	
	{\qquad \qquad \quad}{{\Phi}_{D}^{\bf{T}}}\left( { - {\bf{\Pi }}_D^T{\Phi_A} + {{\bf{F}}^T}{\bf{q}}} \right) = 0\\
	{\qquad \qquad \quad}{{\bf{E}}}{{\Phi_A}} = {{\bf{e}}}\\
	{\qquad \qquad \quad}{{\bf{F}}}{{\Phi_D}} = {{\bf{f}}}\\
	{\qquad \qquad \quad}{\Phi_A} \ge 0\\
	{\qquad \qquad \quad}{\Phi_D} \ge 0
	\end{array}
	\label{eq: MSG_LP}
	\end{equation}
	The value of ${\bf p}, {\bf q}, {\Phi_A}, {\Phi_D}$ that satisfies the constraints in~(\ref{eq: MSG_LP}) can be found through the Linear Complementary Programing (LCP)~\cite{cottle1992linear} by introducing the non-negative vector ${{\bf{z}} = {\left( {\Phi_A,\Phi_D, \bf{p}^{'},\bf{p}^{''}, \bf{q}^{'}}, \bf{q}^{''} \right)^T}}$ where $\bf{p}^{'}, \bf{p}^{''}$ and $\bf{q}^{'}, \bf{q}^{''}$ are non-negative vectors of the same dimension as $\bf{p}=\bf{p}^{'}-\bf{p}^{''}$ and $\bf{q}=\bf{q}^{'}-\bf{q}^{''}$. Furthermore, we let
	\begin{equation}
	\bf{M} = 
	\left[
	\begin{array}{cccccc}
	0 & -\bf{\Pi_A}& \bf{E}^T& -\bf{E}^T & 0 & 0 \\ 
	-\bf{\Pi_D}^T& 0 & 0 & 0& 0 & -\bf{F}^T \\ 
	-\bf{E} & 0 & 0 & 0& 0 & 0 \\ 
	-\bf{E} & 0 & 0 & 0& 0 & 0 \\ 
	0 & -\bf{F} & 0 & 0& 0 & 0 \\ 
	0 & -\bf{F} & 0 & 0& 0 & 0
	\end{array} 	
	\right]
	\end{equation}
	and ${{\bf{b}}^T} = \left( {\begin{array}{*{20}{c}}
		0,&0,&\bf{e},& { - \bf{e}},&\bf{f},&{ - \bf{f}}
		\end{array}} \right)^T$. Then, we have the LCP problem as follows:
	\begin{equation}
	\begin{array}{*{20}{l}}
	{{\rm{find}}\quad {\bf{z}}}\\
	{{\rm{s}}.{\rm{t}}\quad {\bf{Mz + b}} \ge {\bf{0}}}\\
	{\qquad {{\bf{z}}^{\bf{T}}}\left( {{\bf{Mz + b}}} \right) = 0}\\
	{\qquad {\bf{z}} \ge 0}
	\end{array}
	\label{eq:LCP}
	\end{equation}
	The LCP problem above could be solved by the \textit{Lemke algorithm}~\cite{koller1996efficient, von1996tracing, cottle1992linear}. The main idea of the Lemke algorithm is to apply the pivoting operation in the complementary problem. A more detailed description can be found in~\cite{cottle1992linear}. Since a feasible solution at least exists for the formulated LCP problem~\cite{koller1996efficient}, hence the solution of (\ref{eq: MSG_LP}) could be found by considering the solution of (\ref{eq:LCP}). This solution is the NE point of the game. We have the following propositions. 
	
	\begin{proposition}
		In the MSG game, the payoff matrix's size in the sequence-form representation is much smaller than the payoff matrix's size in the strategic-form representation.
		\begin{IEEEproof}
			We observe the payoff matrix in the sequence-form representation is much smaller than one in the strategic-form because $K_4 \ll K_3$. In particular, the payoff matrix in the sequence-form is linear in the size of the game whereas the payoff matrix in the strategic-form is generally exponential. For example, we consider for a CRN with the number of available channels ${N=4}$ where the attack capability of the attacker and the surveillance capability of the defender $M=L=1$. The size of the corresponding payoff matrix is ${86 \times 48}$, which is much smaller than one with the strategic-form representation (${5 \times (5 \times 12^{6})}$).
		\end{IEEEproof}
	\end{proposition}
	\begin{remark}
		Below are some results from Proposition 1.
		\begin{itemize}
			\item The strategy space of the sequence-form representation is exponentially smaller than the strategy space of the strategic-form representation. Since the two methods operate similarly~\cite{koller1996efficient, von1996tracing}, the run time of each algorithm depends on the size of the input. Thus, it is exponentially faster to run the Lemke algorithm on the sequence-form than the L–H algorithm on the strategic-form.
			\item  Since the sequence-form representation is much more compact than the strategic-form representation, we therefore adopt the sequence-form representation approach to determine the NE strategy of the MSG game.
		\end{itemize}
	\end{remark}
	\begin{proposition}
		If $L \geq N$, the NE point of the MSG game is unique and corresponds to the combination result of $N$ independent single-channel surveillance games (results in our previous work~\cite{7103339}), each with one channel in the set of available channels.
		\begin{IEEEproof}
			(\emph{sketch})
			We consider the CRN with $N$ available channels. When ${L \geq N}$m the defender can perform to surveillance on all disallowed channels. In this case, it is equivalent to perform single-channel surveillance game of all disallowed channels independently. For each single-channel surveillance game, the NE point is unique and has been figured out in~\cite{7103339}. Therefore, the NE point of the MSG game with $L \geq N$ is unique and could be determined by combining $N$ unique NE points of $N$ single-channel surveillance games.
		\end{IEEEproof}	
	\end{proposition}
	

\begin{figure*}[!t]
	\centering
	\subfloat[$k_C=0.4$]{\includegraphics[width=2in]{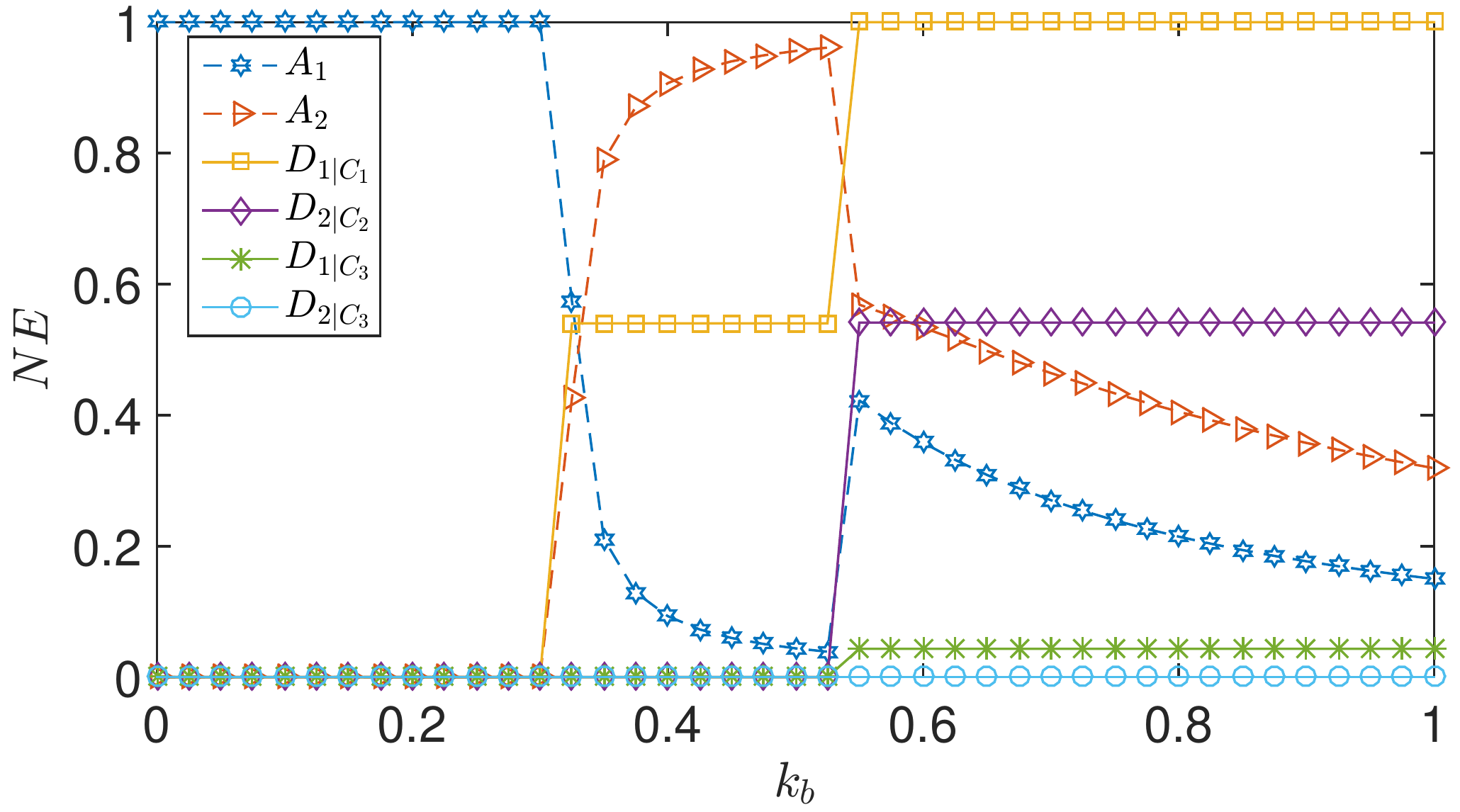}
		\label{fig:2d}}
	\hfil
	\subfloat[$k_C=3$]{\includegraphics[width=2in]{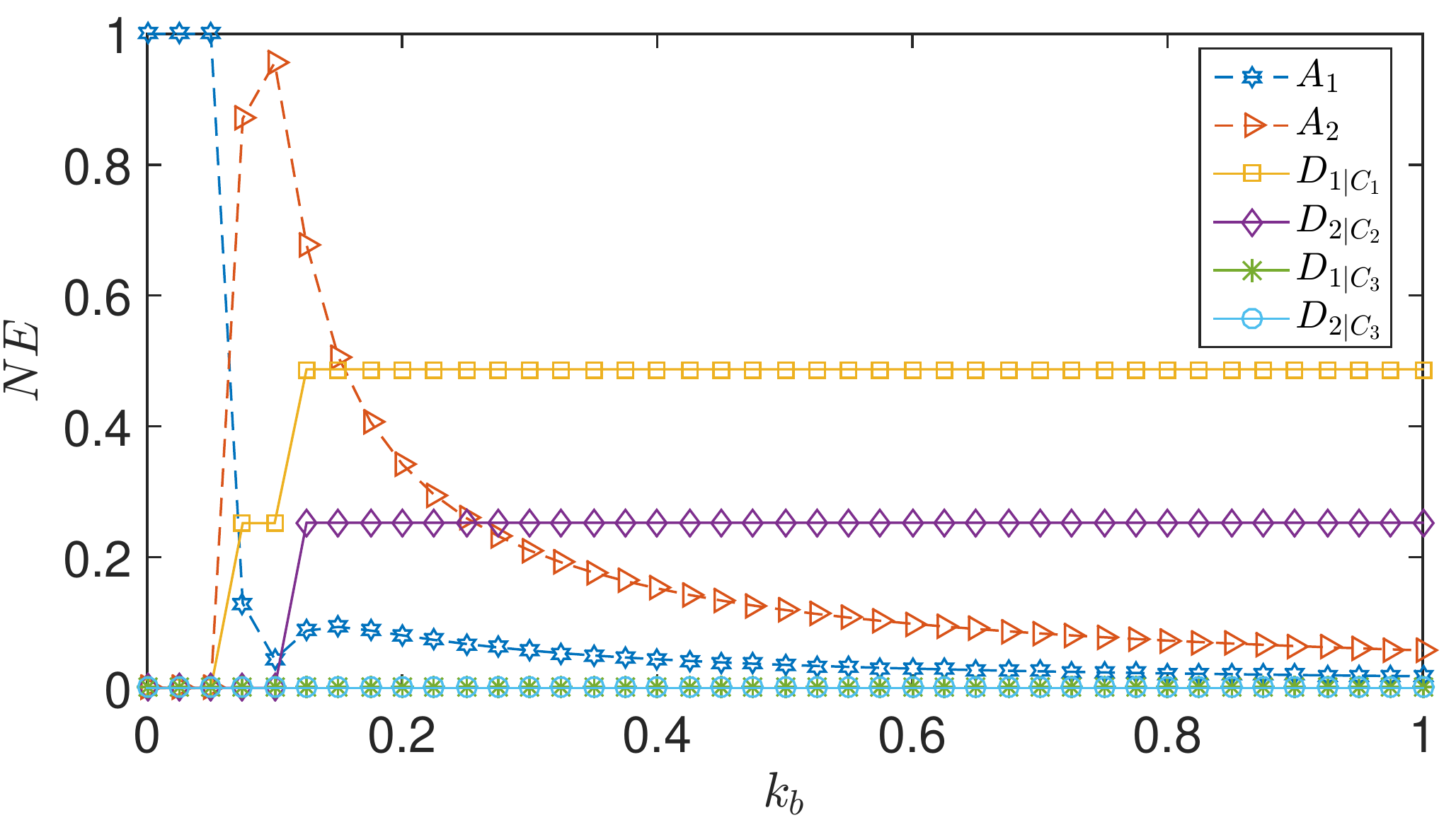}
		\label{fig:2e}}
	\hfil
	\subfloat[$k_C=10$]{\includegraphics[width=2in]{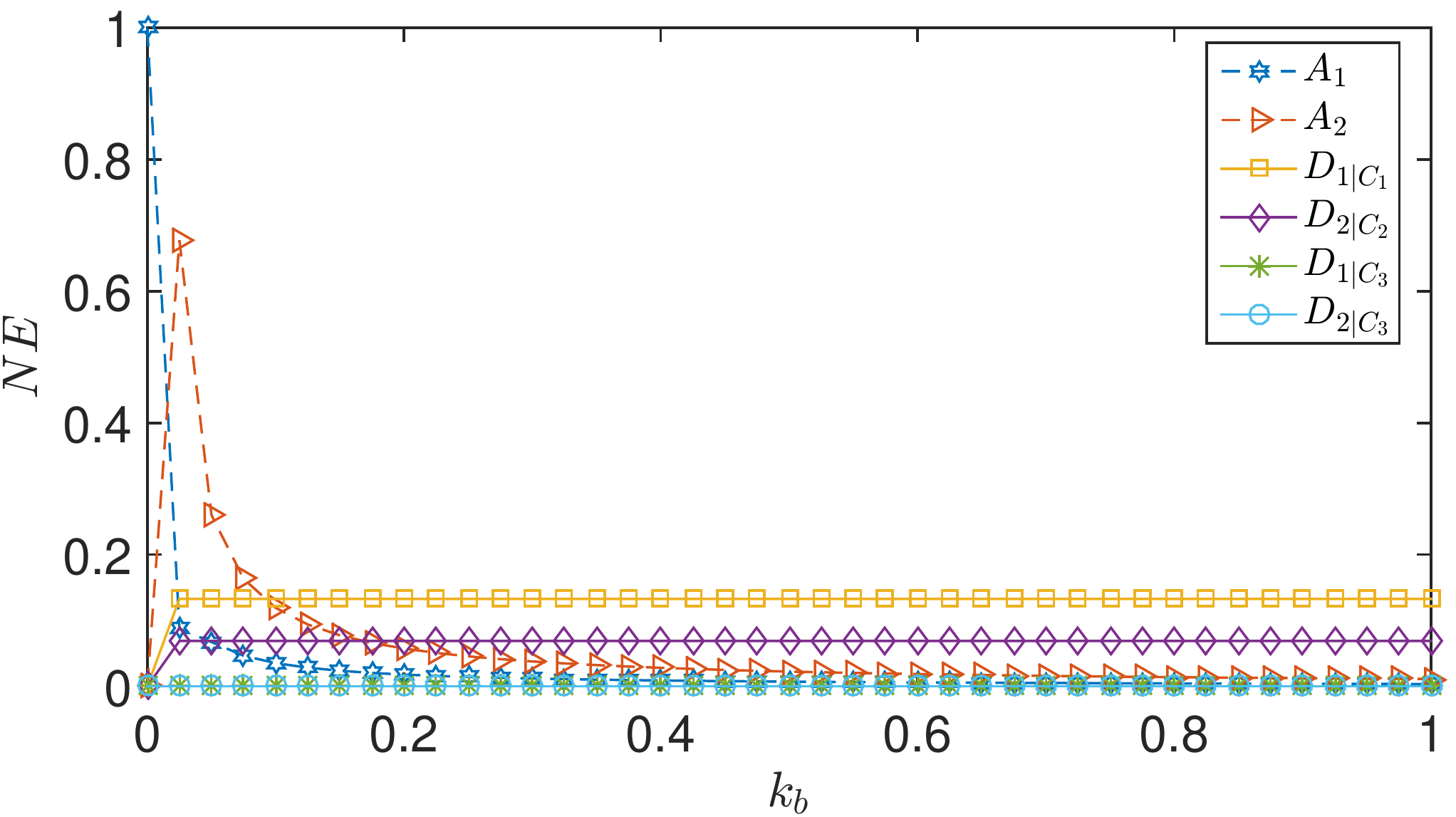}
		\label{fig:2f}}
	\caption{The NE strategy of the MSG game for $N=2$, with channel characteristics $\pi_1=0.2$ and $\pi_2=0.5$.}
	\label{fig: 2channels}
\end{figure*}
	\section{Numerical Results}	
	In order to analyze the impact of system parameters to the NE point of the MSG game, the numerical computations are conducted in Matlab with the Gambit toolbox~\cite{mckelvey2006gambit} for game theory. We also assume the sensing system adopts the energy detection method and the average SNRs of the primary signal received at sensor are $-10$dB. The false alarm probability is $P_f=0.1$ and the number of samples in the energy detector is ${N_{sample}=1500}$. The detection probability ($P_d$) is then computed through the Constant false alarm rate (CFAR) criterion. To focus on the NE strategy of the game, we assume there is no difference between channels in the attacking costs and in the monitoring gains, and only consider the most significant case $G^{i}_A > C^{i}_A $ $ \forall i = 1 \ldots N$ (i.e., the using gain is higher than the attacking cost of a channel).
	
	We introduce the parameters: i) The penalty factor $k_C$: the ratio between the penalty and the using gain $k_C = P^{i} / G^{i}_A$, ii) the network demand $k_b$: the ratio between the surveillance gain and the penalty $k_b= G^{i}_S / P^{i}$. In addition, we denote by $k_A$ the ratio between the attacking cost and the using gain (i.e., $k_A = C^{i}_A / G^{i}_A$), $k_S$ the ratio between the monitoring cost and the using gain (i.e., $k_S= C^{i}_S / G^{i}_A$).
	
	We first consider the CRN with difference cases of channel numbers $N$. Table~\ref{tab: compare} shows the computational time to determine the NE point of the MSG game by using the sequence-form method and the strategic-form method. The numerical programs are conducted on a Dell Precision M6700 laptop with Intel Core i7 CPUs 2.6 GHz. For $N=2$, two methods use the same run time to determine the NE strategy of the game. For $N=3$, however, the sequence method method is much faster than the strategic-form method. For $N>3$, strategic-form method is unable to provide results while the sequence-form method is feasible. We therefore adopt the sequence-form method to determine the NE strategy of the MSG game.
	\begin{table}[h]
		\centering
		\caption{The computation time to determine the NE point.}
		\label{tab: compare}
		\begin{tabular}{c|cccc}
			& $N=2$ & $N=3$ & $N=4$ & $N=5$ \\
			\hline
			Strategic-form & 2 s & 1960.7 s & $\infty$ & $\infty$ \\
			Sequence-form & 2 s & 32.4 s & 11564 s & $\sim$ 12 h
		\end{tabular} 
	\end{table}
	
	Next, for illustrating the effect of parameters on NE strategy of the MSG game, we consider a CRN with two available channels $({N=2})$, the attack capability of attacker ${M=1}$ and the monitoring capability of the defender ${L=1}$. We assume that $k_A=0.2$, $k_S=0.1$, $\pi_1=0.2$ and ${\pi_2=0.5}$.
	
	Fig.~\ref{fig:2d}, Fig.~\ref{fig:2e} and Fig.~\ref{fig:2f} present the NE strategy of the MSG game for low penalty ($k_C=0.4$), medium penalty ($k_C=2$) and high penalty ($k_C=10$), respectively. To provide a better view, we only plot the NE strategies of the attack actions and the surveillance actions. We observe that for each the penalty factors $k_C$, the distribution of NE points is separated into three regions along with the increase of the network demand $k_b$. 
	First, when the network demand $k_b$ is low, the defender does not need to monitor disallowed channels. Consequently, the attacker could perform the attack to the best channel (channel with low operation of PUs). 
	Second, when the network demand $k_b$ is medium, the attacker has a trend of moving the attack from the best channel to the worst channel. The reason of this trend is that the defender performs its surveillance on the best channel due to the increase of its spectrum demand. 
	Third, when the network demand $k_b$ is very high, the defender increase the surveillance rate at the both channels. As a result, the attacker has a corresponding response by adjusting its attacking rate on the both channels. 
	For each region, the strategy of defender depends on the relation between the network demand $k_b$ and the penalty factor $k_C$. For a fixed penalty factor $k_C$, the defender performs monitoring the disallowed channel with a constant probability for each region. 
	In addition, the size of each region also depends on the penalty factor $k_C$ where the last region corresponding with high network demand $k_b$ is enlarged with penalty factor values. The results mean that the NE strategies of the MSG game are affected by both the penalty factor and the network demand. We conclude that in order to reduce the influence of PUEA, the CRNs should set a high penalty.
	
\begin{figure}[!t]
		\centering
		\subfloat[$\Omega^{(NE)}_D$ - $\Omega^{{{Uni}S}}_D$]{\includegraphics[width=1.5in]{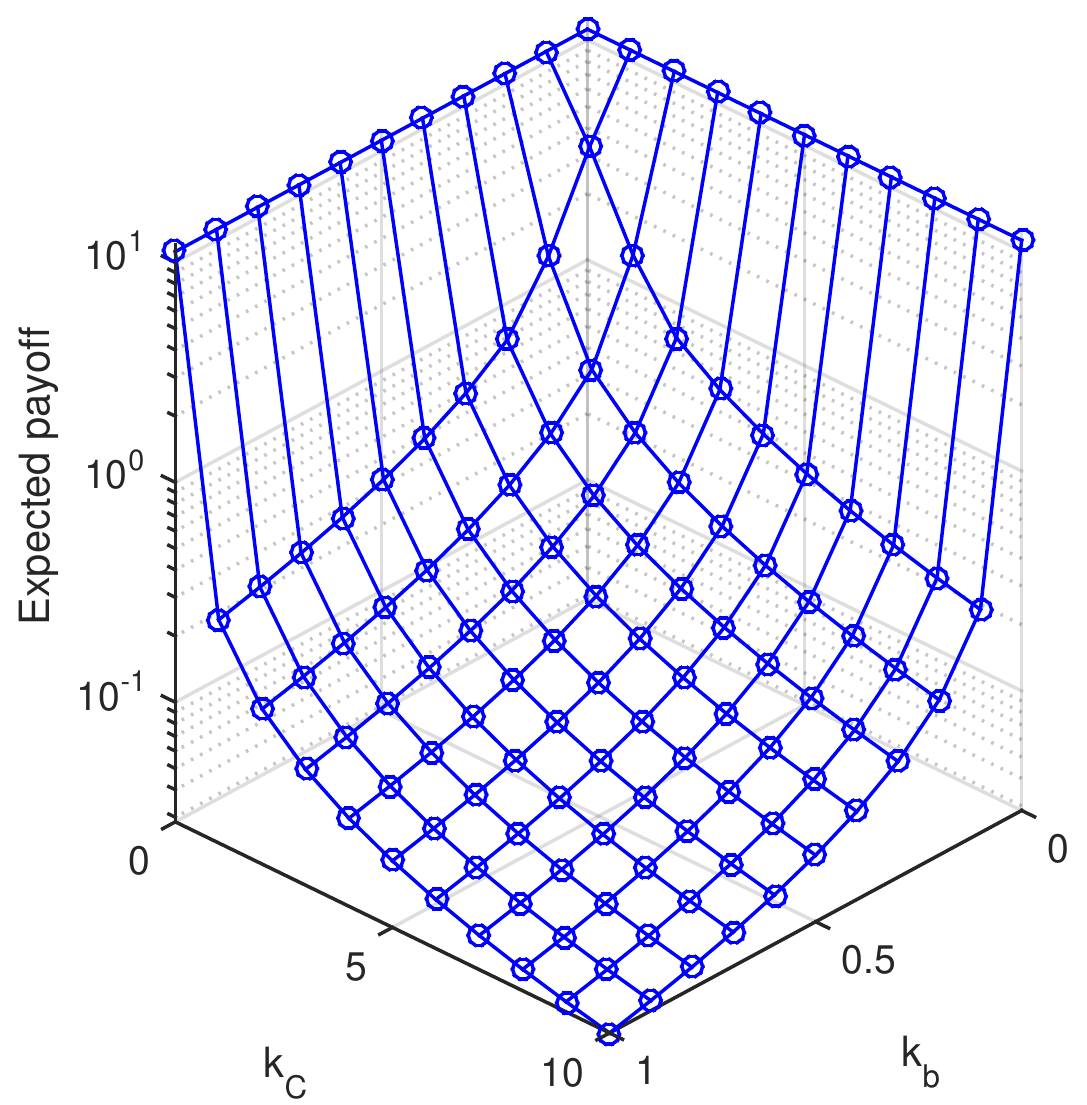}
			\label{fig:3a}}
		\hfil
		\subfloat[$\Omega^{(NE)}_D$ - $\Omega^{(Rand)}_D$]{\includegraphics[width=1.5in]{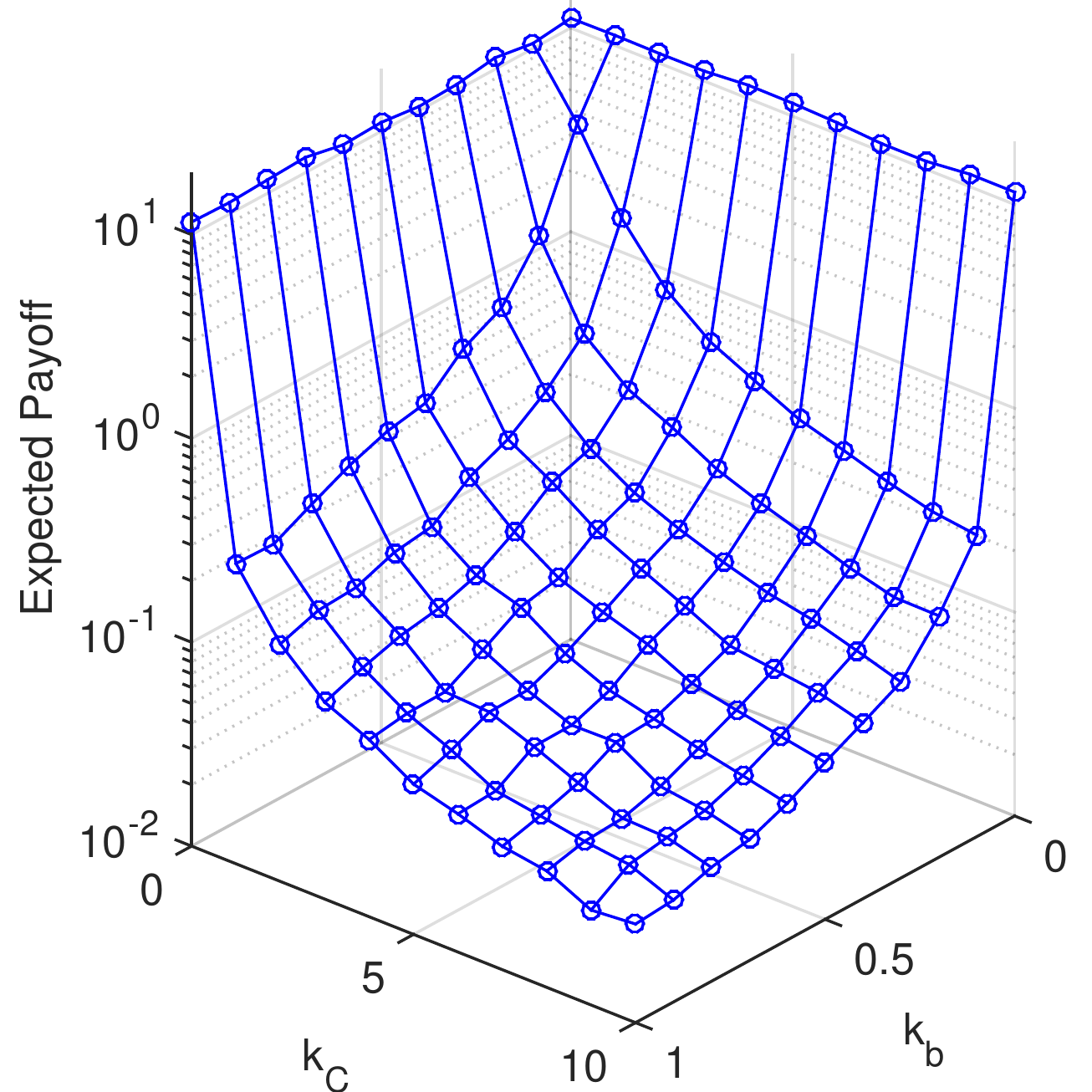}
			\label{fig:3b}}
		\caption{The differences on defender's expected payoffs: (a) between NE and \textit{uniform surveillance} (${Uni}S$) strategies, (b) between NE and \textit{random} (${Rand}$) strategies.}
		\label{fig:NE_efficient}
	\end{figure}

	Figure~\ref{fig:NE_efficient} shows the differences on the defender's expected payoffs when the defender plays (a) NE and \textit{uniform surveillance} strategies, and (b) NE and \textit{random} strategies. The uniform surveillance strategy means that all disallowed channels are uniformly monitored, and the random strategy means that all possible actions are randomly performed.	%

	For the low network demand $k_b$ or the low penalty $k_C$, the monitoring gain is smaller than the monitoring cost. If the defender performs to surveillance the disallowed channels, its expected payoff will be a negative value. For the high penalty factor $k_C$ or the high network demand $k_b$, the NE strategy is the BR for the defender. Therefore, the defender's expected payoff at NE strategy is higher than those at the other strategies. We concluded that the NE strategy is efficient to mitigate the selfish PUEA in multi-channel CRNs.	
\section{Conclusion}
	We have discussed the multi-channel surveillance process to deal with the selfish PUEA in multi-channel cognitive radio networks. 	
	Performing a multi-channel surveillance process on disallowed channels help to identify selfish PUEA attacker. The relation between attack strategies and the surveillance process has been formulated through an extensive-form game. Sequence representation method has been used for obtaining Nash equilibrium. The numerical results have showed the influence of the network demand and the penalty factor on controlling NE strategies and the effectiveness of using NE on mitigating PUEA.  
	Besides, the introduced method has a more efficient computational time compared to the conventional one. We will generalize this method to deal with other PUEA attacker types such as malicious and unknown-attacking type. 
\section*{Acknowledgment}
		This work has been supported by French Ministry of Industry and European CATRENE in CORTIF Project.
	\bibliographystyle{IEEEtran}
	\bibliography{IEEEabrv,SG_Bib}

\begin{thebibliography}{10}
\providecommand{\url}[1]{#1}
\csname url@samestyle\endcsname
\providecommand{\newblock}{\relax}
\providecommand{\bibinfo}[2]{#2}
\providecommand{\BIBentrySTDinterwordspacing}{\spaceskip=0pt\relax}
\providecommand{\BIBentryALTinterwordstretchfactor}{4}
\providecommand{\BIBentryALTinterwordspacing}{\spaceskip=\fontdimen2\font plus
\BIBentryALTinterwordstretchfactor\fontdimen3\font minus
  \fontdimen4\font\relax}
\providecommand{\BIBforeignlanguage}[2]{{%
\expandafter\ifx\csname l@#1\endcsname\relax
\typeout{** WARNING: IEEEtran.bst: No hyphenation pattern has been}%
\typeout{** loaded for the language `#1'. Using the pattern for}%
\typeout{** the default language instead.}%
\else
\language=\csname l@#1\endcsname
\fi
#2}}
\providecommand{\BIBdecl}{\relax}
\BIBdecl

\bibitem{Murty2011}
P.~Murty, ``{SenseLess: A Database-Driven White Spaces Network},'' \emph{IEEE
  Transactions on Mobile Computing}, vol.~11, no.~2, pp. 189--203, Feb 2012.

\bibitem{Ruiliang2008_PUE}
R.~Chen and et~al, ``{Defense against Primary User Emulation Attacks in
  Cognitive Radio Networks},'' \emph{IEEE Journal on Selected Areas in
  Communications}, vol.~26, no.~1, pp. 25--37, Jan 2008.

\bibitem{Ruiliang2008_both}
------, ``{Toward secure distributed spectrum sensing in cognitive radio
  networks},'' \emph{IEEE Communications Magazine}, vol.~46, no.~4, pp. 50--55,
  April 2008.

\bibitem{Mohapatra2013_PUE}
S.~Chen and et~al, ``{Hearing Is Believing: Detecting Wireless Microphone
  Emulation Attacks in White Space},'' \emph{IEEE Transactions on Mobile
  Computing}, vol.~12, no.~3, pp. 401--411, March 2013.

\bibitem{6231141}
Y.~Tan and et~al, ``{Primary user emulation attack in dynamic spectrum access
  networks: a game-theoretic approach},'' \emph{IET Communications}, vol.~6,
  no.~8, pp. 964--973, May 2012.

\bibitem{Li2010_PUE}
H.~Li and Z.~Han, ``{Dogfight in Spectrum: Combating Primary User Emulation
  Attacks in Cognitive Radio Systems, Part I: Known Channel Statistics},''
  \emph{IEEE Transactions on Wireless Communications}, vol.~9, no.~11, pp.
  3566--3577, November 2010.

\bibitem{7103339}
N.~N. Thanh and et~al, ``Surveillance strategies against primary user emulation
  attack in cognitive radio networks,'' \emph{IEEE Transactions on Wireless
  Communications}, vol.~14, no.~9, pp. 4981--4993, Sept 2015.

\bibitem{7194088}
T.~Duc-Tuyen and et~al, ``Extra-sensing game for malicious primary user
  emulator attack in cognitive radio network,'' in \emph{2015 European
  Conference on Networks and Communications (EuCNC)}, June 2015, pp. 306--310.

\bibitem{Felegyhazi2006}
M.~Felegyhazi and J.~Hubaux, ``{Game Theory in Wireless Networks: A
  Tutorial},'' \emph{Computing Surveys, ACM}, 2006.

\bibitem{bwang2010}
B.~Wang and et~al, ``{Game theory for cognitive radio networks: An overview},''
  \emph{Computer Networks}, vol.~54, no.~14, pp. 2537--2561, 2010.

\bibitem{koller1996efficient}
D.~Koller and et~al, ``Efficient computation of equilibria for extensive
  two-person games,'' \emph{Games and Economic Behavior}, vol.~14, no.~2, pp.
  247--259, 1996.

\bibitem{von1996tracing}
B.~von Stengel and et~al, \emph{Tracing equilibria in extensive games by
  complementary pivoting}.\hskip 1em plus 0.5em minus 0.4em\relax Tilburg
  University, 1996.

\bibitem{Harsanyi2004}
J.~Harsanyi, ``{Games with Incomplete Information Played by "Bayesian" Players,
  I-III},'' \emph{Manage. Sci.}, vol.~50, no. 12 Supplement, pp. 1804--1817,
  Dec. 2004.

\bibitem{Lemke1964}
C.~Lemke and J.~{Howson Jr}, ``{Equilibrium Points of Bimatrix Games},''
  \emph{Journal of the Society for Industrial and Applied Mathematics},
  vol.~12, no.~2, pp. 413--423, 1964.

\bibitem{cottle1992linear}
R.~W. Cottle, J.-S. Pang, and R.~E. Stone, \emph{The linear complementarity
  problem}.\hskip 1em plus 0.5em minus 0.4em\relax Siam, 1992, vol.~60.

\bibitem{mckelvey2006gambit}
\BIBentryALTinterwordspacing
R.~D. McKelvey and et~al, ``Gambit: Software tools for game theory,'' 2014.
  [Online]. Available: \url{http://www.gambit-project.org.}
\BIBentrySTDinterwordspacing

\end{thebibliography}
\end{document}